\documentclass[12pt]{iopart}

\usepackage{graphicx}
\usepackage{bm}
\usepackage{amssymb}

\begin{document}
\title[
Critical currents and non-dissipative drag  in quantum Hall
multilayers
 ]{Critical currents and giant non-dissipative drag for
superfluid electron-hole pairs in quantum Hall multilayers}
\author{L Yu Kravchenko, D V Fil}
\address{Institute for Single Crystals, National Academy of
Science of Ukraine, Lenin av. 60, Kharkov 61001, Ukraine}
\eads{\mailto{sotice@mail.ru}, \mailto{fil@isc.kharkov.ua}}

\begin{abstract}
Superfluid properties of electron-hole pairs in a quantum Hall
four-layer system are investigated. The system is considered as a
solid state realization of a two-component superfluid Bose gas
with dipole-dipole interaction. One superfluid component is formed
in the top bilayer and the other component - in the bottom one. We
obtain the dispersion equation for the collective mode spectrum
and compute the critical parameters (the critical interlayer
distance and the critical currents) versus the filling factor. We
find that the critical currents of the components depend on each
other. The maximum critical current of a given component  can be
reached if the current of the other component  is equal to zero.
The non-dissipative drag effect between the components is studied.
It is shown that in the  system considered the drag factor is very
large. Under appropriate conditions it can be about 10 per sent,
that is at least in three order larder than one predicted for
two-component atomic Bose gases.
\end{abstract}


\maketitle

\section{Introduction}
\label{intro}

Among the objects that demonstrate Bose-Einstein condensation or
superconductivity considerable attention is given to two-component
systems. Particularly, beginning from the paper by Andreev and
Bashkin \cite{1} the possibility of a non-dissipative drag between
superfluid (superconducting) components moving with different
velocities was discussed \cite{2,3,4,5,6} (see also the review
\cite{7}). The Andreev-Bashkin effect was also considered in
astrophysics in the context of superfluid models of neutron stars
\cite{8,9}. A related problem - the critical velocities in
two-component superfluid systems was studied in a recent paper
\cite{10}. It was shown that critical velocities are essentially
different in the case when two components move with the same
velocities and in the case when one of the components does not
move. In the latter case the critical velocity of a moving
component can be much higher.

Although two-component superfluid atomic Bose gases have been
realized in laboratories \cite{11}, there are certain problems in
experimental observation of the effects caused by relative motion
of the components. On the one hand, it is not so simple to create
a relative flow of superfluid components in a mixture of Bose
gases. On the other hand, a spatial separation of components takes
place in two-component mixtures confined in a trap. In the gases
with point interaction the spatial separation results in a
disappearance of the effects caused by inter-specie interaction.
To overcome these difficulties one can turn to systems where
components can be kept spatially separated in a controllable way
(that gives a possibility to provide a flow of the components with
different velocities). The interaction between the components in
such systems should contain a  long-range part. For example, one
can deal with a Bose gas with dipole-dipole interaction confined
in a double-layer trap \cite{5}.

In this paper we study a  solid-state system where a two-component
Bose gas with the dipole-dipole interaction can be realized. We
consider a multilayer electron system where electrons from one
layer couple with holes from the adjacent layer. Since such Bose
particles (electron-hole pairs) have a small mass they may
demonstrate superfluid behavior at rather high temperatures (much
higher than required for the Bose-Einstein condensation of alkali
metal vapors). Electron-hole pairs in such systems have large
dipole momentum and  the dipole-dipole interaction determines, in
the main part, collective properties of a gas of such pairs.

\begin{figure}
\begin{center}
\includegraphics[width=10cm]{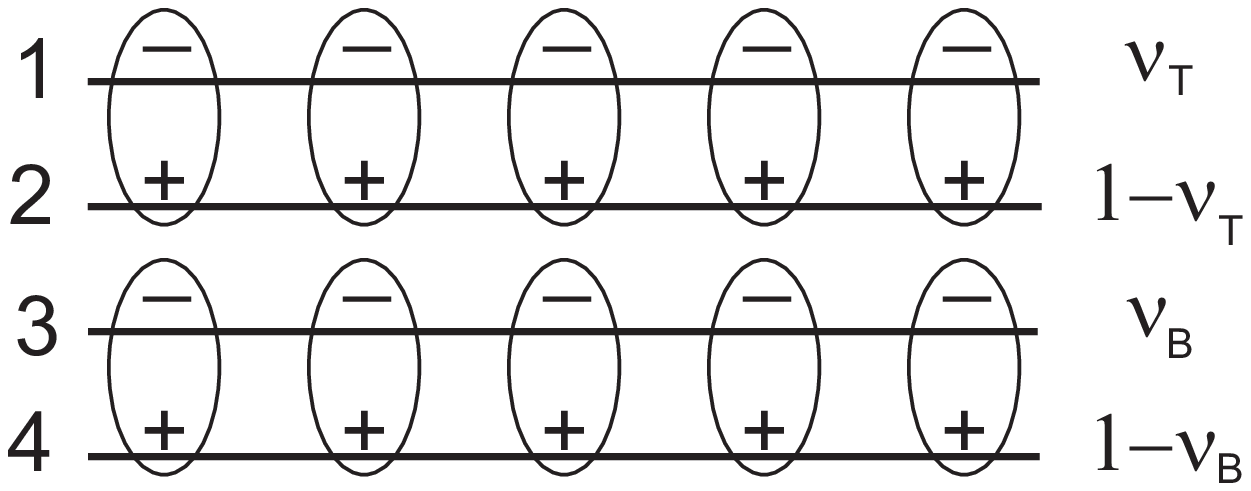}
\end{center}
\caption{The four-layer system with two-component gas of
electron-hole pairs.} \label{figure1}
\end{figure}

To be more specific, we consider a four-layer electron system in a
strong perpendicular to the layers magnetic field (a multilayer
quantum Hall system).  For bilayer quantum Hall systems with total
filling factor equal to unity the theory predicts \cite{16} the
existence of a superfluid condensate of indirect excitons in the
systems. An indirect exciton corresponds to a bound state of an
electron belonging to one layer and a hole (an empty state in the
lowest Landau level) belonging to the other layer. This prediction
was partially confirmed in experiments \cite{17,e1,e2}.
Bose-Einstein condensation of metastable (optically generated)
indirect excitons in zero magnetic field was also observed
\cite{e3,e4}. As was shown in \cite{18}, multi-component excitonic
superfluid condensates can be realized in multilayer quantum Hall
systems with even number of the layers and the average filling
factor per layer equal to one-half. According to \cite{18},
electron-hole pairs emerge in separate bilayers, i.e., a given
excitonic component belongs to a given double-layer complex. Here
we consider a four-layer system with the filling factors of the
layers $ \nu_1 =\nu_\mathrm{T} $, $ \nu_2=1-\nu_\mathrm{T} $, $
\nu_3 =\nu_\mathrm{B} $, $ \nu_4=1-\nu_\mathrm{B} $. In such a
system one specie is formed by coupled electrons and holes from
the layers 1 and 2, while the other specie - by coupled carriers
from the layers 3 and 4 (Figure \ref{figure1}).

In the system considered a flow of electron-hole pairs is
equivalent to two oppositely directed electrical currents in the
adjacent layers. Therefore, a superfluid state of such pairs can
be considered as a specific superconducting state. Counterflow
supercurrents can carry an electrical current from the source
situated at one end of the system to the load situated at the
opposite end (if the interlayer tunnelling is negligible small,
the dissipation is negligible, as well \cite{20a}). Experimentally
one can provide separate contacts to each layer. It allows to
control and measure the supercurrents in each bilayer complex.
Therefore, it is more appropriate to formulate the problems not in
terms of superfluid velocities, but in terms of supercurrents.

In this paper we address two problems. First, we investigate
critical currents in a four-layer quantum Hall system. We show
that the behavior of critical currents is qualitative the same as
of critical velocities in two-component Bose gases (with the
important advantage that the effect in multilayers can be
registered by electrical measurements). Then, we consider the
non-dissipative drag between the components and compute the drag
factor. We predict very large drag factor for quantum Hall
multilayers: under appropriate conditions it can reach  10 per
sent (in three order higher than the most optimistic estimates for
atomic Bose gases).

Our study is based on the analysis of collective mode spectra. We
follow the approach proposed in \cite{20} for the study  of
bilayer systems with zero imbalance of filling factors
($\nu_1=\nu_2=1/2$). In Sec.\ref{s2} we extend the approach
\cite{20} for the case of an arbitrary imbalance. In Sec.\ref{s3}
we obtain the spectra of excitations for the four-layer system,
find the critical interlayer distance versus the filling factor,
and obtain the relation between the critical currents. The
non-dissipative drag effect is considered in Sec.\ref{s4}.

\section{The approach}
\label{s2}

Let us begin with the discussion of a mechanism that determines
critical supercurrents in quantum Hall bilayers (multilayers). In
bulk superconductors the restriction on the value of the
supercurrent emerges from the requirement that the magnetic field
produced by electrical currents should be lower than the
thermodynamic critical magnetic field. For thin films the critical
magnetic field is higher than for bulk superconductors, and at
small thickness $w$ of the film it increases by the law
$H_\mathrm{c}\propto 1/w$. Due to almost two-dimensional character
of conducting layers in quantum Hall systems the critical magnetic
field should be high. The critical current is determined by an
essentially different mechanism. This mechanism is just the
generalization of the Landau mechanism of destruction of
superfluidity (governed by the Landau criterion). The critical
current can be found from the requirement that in a
superconducting state the energies of all collective excitations
being real valued and positive. Going ahead we note that
electrical currents close to critical ones produce magnetic fields
much smaller than the terrestrial magnetic field. And, indeed, the
magnetic mechanism of destruction of superconductivity is
irrelevant for the quantum Hall bilayers (multilayers).

For more transparency, we will describe the approach with
reference to a bilayer system. Let a double-layer electron system
is situated in a perpendicular to the layers magnetic field $B$.
The electron density $\rho=\rho_1+\rho_2$ satisfies the condition
$\nu_\mathrm{tot}=2 \pi l^2 \rho=1$, where $ l =\sqrt {\hbar c/e
B} $ is the magnetic length, i.e. the  filling factor  of the
layer 1 is $ \nu_1 =\nu $ ($ \nu <1$), and the filling factor of
the layer 2 is $ \nu_2=1-\nu $. The value $ \nu $ is the parameter
of the problem.  Since $\nu_1,\nu_2<1$ all the carriers belong to
the lowest Landau level (under assumption that the Coulomb energy
is small compared to the energy gap between the Landau levels). We
take the Hamiltonian in the lowest Landau level approximation:
\begin{equation} \label{5}
H = \frac {1} {2S} \sum _ {n, n ' =1} ^2 \sum_\mathbf {q} V _ {n,
n '} (q) \left \{\rho_n (\mathbf {q}) \rho _ {n '} (-\mathbf {q})
- \delta _ {n, n '} \rho_n (0) \exp\left (-\frac {l^2 q^2} {2}
\right) \right \},
\end{equation}
where $S=L_\mathrm{x} L_\mathrm{y} $ is the area of the layer,
$V_{n, n'}(q) = {2\pi e^2} \rme^{-d q |n-n'|}/ \varepsilon q$ is
the Fourier-component of the Coulomb potential, $d$ is the
distance between the layers, $\varepsilon$ is the dielectric
constant, and
\begin{equation} \label{5a}
\rho_n (q) = \sum_X c_n ^ + (X + \frac {q_\mathrm{y} l^2} {2}) c_n
(X - \frac {q_\mathrm{y} l^2} {2}) \exp ({\rmi} q_\mathrm{x} X -
\frac {q^2 l^2} {4})
\end{equation}
is the Fourier-component of the electron density in the $n$-th
layer. In (\ref{5a}) $c_n ^ + (X) $ and $c_n (X) $ are the
creation and annihilation operators for electrons in  the $n$-th
layer in the state described by the wave function $\psi_X (\mathbf
{r}) = \exp \left (- {\rmi} X y/l^2-(x-X) ^2/{2 l^2} \right)$. We
imply that the interlayer tunnelling amplitude $t$ is much smaller
than the Coulomb energy $E_c= e^2/\varepsilon l$ and neglect the
tunnelling in the Hamiltonian (\ref{5}). We will discuss the
validity of such an approximation in more details at the end of
this section. Here we just mention that in the bilayer systems
used in experiments the tunneling amplitude $t\approx 50$ $\mu$K
\cite{17} that is in 6 order smaller than the Coulomb energy.

The state with electron-hole pairing can be described by a
BCS-like many-body wave function $
    | \Psi\rangle =\prod_X\left [u_X+v_X h_1 ^ + (X) c_2 ^ + (X) \right]
    |vac\rangle $,
where $h_1 ^ + $ is the creation operator of the hole in the layer
1 and $ |vac\rangle $ is the vacuum state defined as a state with
completely filled layer 1 and empty layer 2. The $u-v $
coefficients  satisfy the condition $ |u_X | ^ 2 + | v_X | ^ 2=1$.
This function can be presented in another equivalent form
\begin{equation} \label{vf1}
    | \Psi\rangle =\prod_X\left [\cos\frac {\theta _ {X}} {2} c_1 ^ + (X)
    + \rme ^ {{\rmi}\varphi_X} \sin\frac {\theta_X} {2} c_2 ^ + (X)
    \right] |0\rangle,
\end{equation}
where $\theta_X=\theta(X)$ and $\varphi_X=\varphi(X)$ are
arbitrary functions. The quantity $\theta$ can be connected with
the local filling factors  $\nu_{X,1(2)}=(1\pm \cos \theta_X)/2$.
One can see that $\varphi_X$ is the phase of the order parameter
$\Delta_X=\langle\Psi |\, c ^ + _ {1 X} c _ {2 X} |
\Psi\rangle=\rme^{{\rmi}\varphi_X}\sqrt{\nu_X(1-\nu_X)}$ which
corresponds to the electron-hole pairing. At $\theta={\rm const}$
and $\varphi=0$  the function (\ref{vf1}) in the coordinate
representation coincides with the famous (1,1,1) Halperin wave
function (see, for instance, \cite{21}).

In the state (\ref {vf1}) the energy of the system reads as
\begin{eqnarray} \label{10}
 E = \frac {1} {2 L_\mathrm{y}} \sum _ {X, X '} \Big \{\left [H
(X-X ') - F_\mathrm{S} (X-X ') \right] \cos \theta_X \cos \theta _
{X '} \cr
 - F_\mathrm{D} (X-X ') \sin\theta_X \sin\theta _ {X '}
\cos (\varphi_X - \varphi _ {X '}) \Big \},
\end{eqnarray}
where the quantities
\begin{eqnarray} \label{11}
H (X) = \frac {e^2} {2 \varepsilon} \int _ {-\infty} ^ {\infty}
\rmd q \frac {1-\rme ^ {-|q | d}} {|\, q |} \; \rme ^ {{\rmi} q X
- \frac {q^2 l^2} {2}},\cr F_\mathrm{S} (X) = \frac {e^2} {2
\varepsilon} \; \rme ^ {-\frac {X^2} {2 l^2}} \int _ {-\infty} ^
{\infty} \frac {\rmd q} {\sqrt {q^2 + X^2/l^4}} \; \rme ^ {- \frac
{q^2 l^2} {2}},\cr F_\mathrm{D} (X) = \frac {e^2} {2 \varepsilon}
\; \rme ^ {-\frac {X^2} {2 l^2}} \int _ {-\infty} ^ {\infty} \frac
{\rmd q} {\sqrt {q^2 + X^2/l^4}} \; \rme ^ {-| q | d - \frac {q^2
l^2} {2}}
\end{eqnarray}
describe the direct Coulomb interaction, the exchange interaction
in a given layer, and the exchange interaction between the layers,
respectively.

We consider  excitations above a homogeneous state with a
stationary superflow of electron-hole pairs along the $x$
direction. Such a state corresponds to $ \theta_X =\theta_0$
independent of $X$ and the phase $ \varphi_X=QX $ linear in $X$.
The energy of the homogeneous state is found from Eq.(\ref{10})
and reads as
\begin{equation} \label{6-1}
    E ^ {(0)} = \frac {S} {4 \pi l^2} \Bigg(\left [\mathcal {H} (0) - \mathcal {F} _\mathrm{S}
(0) \right] \cos^2 \theta_0 - \mathcal {F} _\mathrm{D} (Q)
\sin^2\theta_0 \Bigg).
\end{equation}
Here the functions displayed calligraphically indicate the
Fourier-transforms defined as $\displaystyle \mathcal {A} (q) =
(1/2 \pi l^2) \int _ {-\infty} ^ {\infty} {\rmd X} \exp (-{\rmi} q
X) A (X) $. The explicit expressions for the quantities in
(\ref{6-1}) are
\begin{equation} \label{62}
    \mathcal {H} (q) = \frac {e^2} {2 \varepsilon l^2} \; \rme ^ {-\frac {q^2 l^2} {2}}
    \frac {1 - \rme ^ {-d |q |}} {|\, q |},
\end{equation}
\begin{equation} \label{63}
    \mathcal {F} _\mathrm{S} (q) = \frac {e^2} {2 \varepsilon} \int_0 ^ {\infty}
    \rmd k \, \rme ^ {-\frac {k^2
    l^2} {2}} J_0 (k q l^2), \quad
\end{equation}
\begin{equation} \label{64}
    \mathcal {F} _\mathrm{D} (q) = \frac {e^2} {2\varepsilon} \int_0 ^ {\infty}
    \rmd k \; \rme ^ {-\frac {k^2
    l^2} {2}} J_0 (k q l^2) \; \rme ^ {-k d}.
\end{equation}

Fluctuations over the stationary state can be parametrized as
$\tilde {m} _z (X)= \cos \theta_X-\cos \theta_0$ and  $
\tilde{\varphi}_X=\varphi_X -Q X$.  The energy of  fluctuations in
quadratic approximation has the form
\begin{equation} \label{32}
\fl E_{{\rm fl}}= \frac {S} {4 \pi l^2} \sum_q \Big [\tilde{m} _
{z} (-q) \mathcal {K} _ {zz} (q) \tilde{m} _ {z} (q) + 2
\tilde{m}_z (-q) \mathcal {K} _ {z \varphi} (q) \tilde{\varphi}
(q)  + \tilde{\varphi} (-q) \mathcal {K} _ {\varphi \varphi} (q)
\tilde{\varphi} (q) \Big].
\end{equation}
In (\ref{32}) the Fourier-components of the fields $\tilde {m} _z
(X)$ and $\tilde{\varphi}_X$ are defined as
\begin{equation} \label{22}
\tilde{m} _ {z} (q) = \frac {2 \pi l^2} {S} \sum_X \tilde{m} _ {z}
(X) \rme ^ {-{\rmi} q X}, \quad \tilde{\varphi} (q) = \frac {2 \pi
l^2} {S} \sum_X \tilde{\varphi} (X) \rme ^ {-{\rmi} q X}.
\end{equation}
In Eq. (\ref{32}) the components of the matrix $\mathcal {K}$ read
as
\begin{equation} \label{26}
\fl \mathcal {K} _ {zz} (q) = \mathcal {H} (q) - \mathcal {F}
_\mathrm{S} (q) + \mathcal {F} _\mathrm{D} (Q)  + \left (\mathcal
{F} _\mathrm{D} (Q) - \frac {\mathcal {F} _\mathrm{D} (q+Q) +
\mathcal {F} _\mathrm{D} (q - Q)} {2} \right) \cot^2 \theta_0,
\end{equation}
\begin{equation} \label{25}
\mathcal {K} _ {z \varphi} = -  {{\rmi}\cos \theta_0}  \frac
{\mathcal {F} _\mathrm{D} (q+Q) - \mathcal {F} _\mathrm{D} (q -
Q)} {2},
\end{equation}
\begin{equation} \label{24}
\mathcal {K} _ {\varphi \varphi} (q) =  {\sin^2 \theta_0}  \left
[\mathcal {F} _\mathrm{D} (Q) - \frac {\mathcal {F} _\mathrm{D}
(q+Q) + \mathcal {F} _\mathrm{D} (q-Q)} {2} \right].
\end{equation}

One can note that Eq. (\ref{26}) diverges at $\theta_0=0$ and
$\theta_0=\pi$ and the approximation (\ref{32}) violates. But such
$\theta_0$ correspond to filing factors $\nu_1=0$ and $\nu_2=1$
(or vice versa). At such filling factor the density of
electron-hole pairs is equal to zero and one cannot speak about
the spectrum of collective excitation in the gas of the pairs. The
cases of $\theta_0$ close to $0$ or $\pi$ corresponds to low
density of the pairs and the approximation (\ref{32}) is valid
under condition that the density fluctuation are small as compared
to the ground state density. The latter condition is equivalent to
$|\tilde{m} _ {z} (q)|\ll \sin \theta_0$.

To quantize the energy (\ref{32}) one notes that $\tilde {m} _z $
and $ \tilde{\varphi} $ are the conjugated quantities and the
commutator of the operators that correspond to these variables is
equal to
\begin{equation} \label{35}
[\hat {m} _ {z} (q), \hat {\varphi} (q ')] = - 2 {\rmi} \frac {2
\pi l^2} {S} \delta _ {q,-q '}.
\end{equation}
The operators $\hat{m} _ {z} (q) $ and $ \hat{\varphi} (q) $ can
be expressed in terms of Bose creation and annihilation operators
in a common way
\begin{equation} \label{37}
\hat {m} _ {z} (q) = A ( {b} _ {q} +  {b} ^ + _ {-q}), \ \hat
{\varphi} (q) = {\rmi} B ( {b} _ {q} -  {b} ^ + _ {-q}),
\end{equation}
where the amplitudes  satisfy the condition $ \displaystyle A B =2
\pi l^2 / S $. Replacing the variables $ \tilde{m} _ {z} (q) $ and
$ \tilde{\varphi} (q) $ in (\ref{32}) with the operators
(\ref{37}) and requiring vanishing of the terms containing two
creation (two annihilation) operators one finds the explicit
expressions for the amplitudes
\begin{equation} \label{37-1}
    A =\sqrt {\frac {2\pi l^2} {S}} \left (\frac {\mathcal {K} _ {\varphi
\varphi} (q)} {\mathcal {K} _ {zz} (q)} \right) ^ {\frac {1} {4}},
\quad
 B =\sqrt {\frac {2\pi l^2} {S}} \left (\frac {\mathcal {K} _ {zz} (q)} {\mathcal {K} _ {\varphi
\varphi} (q)} \right) ^ {\frac {1} {4}}.
\end{equation}
As a result one obtains the Hamiltonian for the collective
excitations
\begin{equation} \label{33}
H_{\mathrm{fl}} =  \sum _ {q} E (q) \left (b ^ + (q) b (q) + \frac
{1} {2} \right),
\end{equation}
where
\begin{equation} \label{39}
E (q) = 2 \left (\sqrt { \mathcal {K} _ {\varphi \varphi}
(q)\mathcal {K} _ {zz} (q)} + \tilde {\mathcal {K}} _ {z \varphi}
(q) \right)
\end{equation}
(with $ \tilde {\mathcal {K}} _ {z\varphi} (q) = {\rm i}  \mathcal
{K} _ {z \varphi} (q) $) is the spectrum of collective
excitations.


It is instructive to compare the spectrum (\ref{39})  with the
Bogolyubov spectrum. Let us introduce formal notations $\epsilon$,
$v$, $\gamma$ and $n$ and present Eq. (\ref{39}) in the Bogolyubov
form
\begin{equation} \label{39-1}
E (q) = \sqrt {\epsilon  (\epsilon   +2 \gamma n )} + \hbar \, q
v.
\end{equation}
 In (\ref{39-1}) the kinetic energy is defined as
\begin{equation} \label{40-1} \epsilon =2\mathcal {F} _\mathrm{D}
(Q) - \mathcal {F} _\mathrm{D} (q+Q) - \mathcal {F} _\mathrm{D}
(q-Q).
\end{equation}
In the long wave limit ($q, Q\ll l ^ {-1} $) this quantity is
reduced to the standard expression for the kinetic energy
$\epsilon= \hbar^2 q^2/2M$, where
\begin{equation} \label{12}
 M = \frac {2 \varepsilon \hbar^2} {e^2 l}
 \left [\sqrt {\frac {\pi} {2}}
 \left (1 + \frac {d^2} {l^2} \right)
 \exp\left (\frac {d^2} {2 l^2} \right)
 \mathrm {erfc} \left (\frac {d} {\sqrt {2} l} \right) -
 \frac {d} {l} \right] ^ {-1}
\end{equation}
is the magnetic mass of the pair (see, for instance, \cite{23}).

\begin{figure}
\begin{center}
\includegraphics[width=10cm]{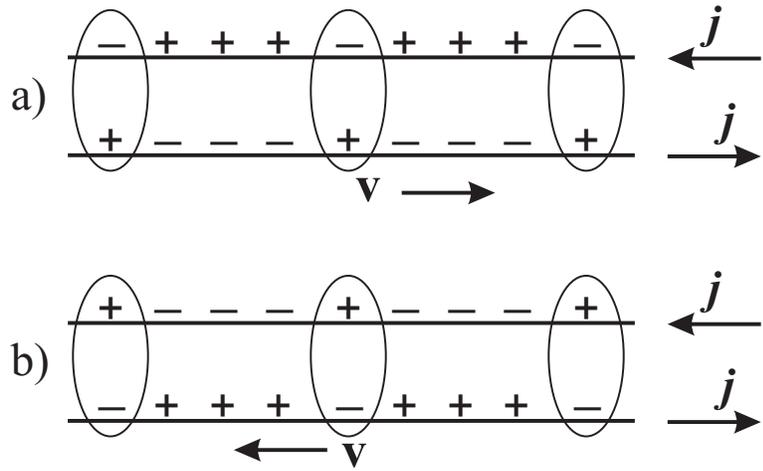}
\end{center}
\caption{Relative directions of electrical currents in the layers
($j$) and the velocities of electron-hole pairs ({\bf v}) at
$\nu<1/2$ (a) and $\nu>1/2$ (b). } \label{figure2}
\end{figure}

In (\ref{39-1}) the density is defined as $n=\nu(1-\nu)/2\pi l^2$.
In the limit $\nu\to 0$ (or $(1-\nu)\to 0$) this quantity
coincides with the density of electron-hole pairs (in this case
one can easily mark out the pairs from the background, see Figure
\ref{figure2}). The factor $\nu(1-\nu)$ appears due to the
electron-hole symmetry of the problem.

The superfluid velocity in (\ref{39-1}) is defined as
\begin{equation} \label{40-3}
v = \frac {\mathcal {F} _\mathrm{D} (q+Q) - \mathcal {F}
_\mathrm{D} (q - Q)} {\hbar q} (2 \nu -1).
\end{equation}
At $q, Q\ll l ^ {-1} $ it reduces to the expression $v=(\hbar \,
Q/M)(1-2\nu)$ that differs from the common expression for the
superfluid velocity by the factor $1-2\nu$. At small $\nu$ this
factor approaches unity and the difference disappears, but at zero
imbalance ($\nu=1/2$) it is equal to zero and the last term in
(\ref{39-1}) vanishes. This feature can be understood from the
following arguments. The sign of $Q$ (gradient of the phase)
determines the direction of the current. If one describes the
supercurrent as motion of electron-hole pairs then the direction
of the current depends on the direction of the superfluid velocity
and on the direction of polarization of electron-hole pairs
(Figure \ref{figure2}). Therefore at given $Q$ the filling factors
$\nu<1/2$ and $\nu>1/2$  corresponds to opposite directions of
superfluid velocities. The factor $(1-2\nu)$ changes its sign
under substitution $\nu\to 1-\nu$  and its appearance in the
expression for $v$ reflects to the electron-hole symmetry of the
problem.

The interaction parameter $\gamma$ in (\ref{39-1}) is given by the
expression
\begin{equation} \label{40-2}
\gamma = 8\pi l^2 \left [\mathcal {H} (q) - \mathcal {F}
_\mathrm{S} (q) + \frac{\mathcal {F} _\mathrm{D} (q+Q)+ \mathcal
{F} _\mathrm{D} (q-Q)}{2}\right].
\end{equation}
The first term in (\ref{40-2}) is caused by the direct Coulomb
interaction between dipoles, two other terms correspond to the
exchange interaction. One can see that at small $q$  the direct
interaction term reduces to $\gamma_0 =4\pi e^2 d/\varepsilon$. It
is just the interaction parameter for a two-dimensional gas of
classical dipoles (polarized perpendicular to the layer). We also
note that in the limit $d\to 0$ (in which the pairs do not
interact with each other) the quantity (\ref{40-2}) approaches
zero, and the collective excitation spectrum (\ref{39-1}) turns to
the spectrum of free particles $E(q)=\epsilon=\hbar^2q^2/2M$ (at
$Q=0$ and $ql\ll 1$).

Thus, the spectrum (\ref{39}) is similar to the Bogolyubov
spectrum but there are certain differences caused by the
electron-hole symmetry. The answer (\ref{39}) at $\nu=1/2$
coincides with the spectrum obtained in \cite{20}. In the low
density limit $\nu\ll 1$ (low concentration of the pairs) and at
small $q$ the result (\ref{39}) reduces to one found in \cite{24}
on the base of the Gross-Pitaevskii equation.

According to the Landau criterium the energy (\ref{39}) should be
positive and real-valued for all $q$.  At $d>d_\mathrm{c}$
(critical $d_\mathrm{c}$ depends on $\nu$) the spectrum becomes
complex-valued already at $Q=0$. It means that at given $d$ and
$\nu$ the superfluid state cannot be realized at all. At
$d<d_\mathrm{c}$ the Landau condition violates at
$Q>Q_\mathrm{c}$. The value $Q_\mathrm{c}$ (that depends on $\nu$
and $d/l$) determines the critical current. The relation between
$Q_\mathrm{c}$ and the critical current can be obtained as
follows.
 At nonzero vector potential the phase gradient $Q$ in the energy (\ref{6-1})
should be replaced with the gauge invariant quantity
\begin{equation} \label{g1}
Q\to Q - (e/\hbar \, c) ({\bf A}_1 - {\bf A}_2),
\end{equation}
where $ {\bf A} _i $ is a vector potential in the $i $-th layer.
In a given layer the current (density of the current) is obtained
from the equation $ j_n=c \, \rmd \epsilon/\rmd A_n $, where $
\epsilon=E^{(0)}/S $ is the energy per unit area. Taking into
account (\ref{g1}), one finds $j_1=-j_2=-(e/\hbar S)\, \rmd
E^{(0)}/\rmd Q$, or, in the explicit form,
\begin{equation} \label{cur}
    j_1 =-j_2 = \frac {e} {\hbar} \frac {1} {4\pi l^2} \sin^2 \theta_0 \frac {\rmd
    \mathcal {F} _\mathrm{D} (Q)} {\rmd Q}.
\end{equation}
Substituting Eq.(\ref{63}) into (\ref{cur}) one obtains the
following relation between $Q$ and the current
\begin{equation} \label{cur1}
    j_1 =-j_2 = - \frac {e^3} {2\pi \varepsilon l^2 \hbar} \,  \nu
    (1-\nu)
\int_0 ^ {\infty} \rmd \tilde {k}\, \rme ^ {-\frac {\tilde {k} ^2
    } {2}} \tilde {k} J_1 (\tilde {k} Q l)\, \rme ^ {-\tilde {k} d/l}.
\end{equation}
At small $Q$ ($Q\ll l$) the current is proportional to the phase
gradient:  $j_1 = (e/\hbar) \rho_\mathrm{s}Q$, where
$\rho_\mathrm{s} = \hbar^2 n/M$ is the superfluid stiffness.

One can easy check  that the value of integral in (\ref{cur1}) at
any $Q$ and $d$ does not exceed 0.45.  Therefore, in any case the
supercurrent is less than $j_\mathrm{m}\approx 0.018 e^3/
\varepsilon l^2 \hbar $. At typical parameters $
\varepsilon=12.5$, $l=100 \AA $ one evaluates $j_\mathrm{m}
\approx 5 $ A/m. Note this value of counterflow currents
corresponds to very small value of parallel to the layers
component of magnetic field produced by the current
($B_\mathrm{y}\approx 6 \cdot 10 ^ {-6} $ T). Under accounting the
Landau criterium this quantity is even smaller.

Now let us discuss the influence of tunnelling on the critical
current. In general case the tunnelling reduces the critical
current, moreover it may destroy the superfluid state completely
\cite{20}. To explain this effect it is instructive to introduce
the Josephson length $\lambda=l \sqrt{2\pi
\rho_s/t\sqrt{\nu(1-\nu)}}$ (see, for instance, \cite{20a,20}).
This quantity determines the size of the soliton (Josephson
vortex) in the bilayer quantum Hall system. If the Josephson
length  is small $\lambda < 2\pi/Q_c$ and the gradient of the
phase caused by the soliton is large (larger than the critical
gradient $Q_c$)  the tunnelling influences significantly on the
critical parameters. But such $\lambda$ correspond to quite large
tunnelling amplitudes ($t/E_C> 10^{-2}$ for $\nu=1/2$ and
$d\approx l$). Thus, in a common experimental situation
($t/E_C\sim 10^{-6}$) the tunnelling can be neglected almost for
all relevant values of $\nu$ and $d/l$.


\section{The excitation spectrum  and critical parameters for the four-layer system }
\label{s3}

Let us turn to the consideration of the four-layer quantum Hall
system where a two-component superfluid gas of electron-hole pairs
can emerge. For definiteness, we  consider the system  with equal
distances between the adjacent layers.
We specify the case where
supercurrents in both components are directed along the same axis.

The Hamiltonian of the system has the form  (\ref{5}) (with the
summation over four layers). According to the result of Refs.
\cite{18} the electron-hole pairs are formed separately in  the
$n=1,2$ bilayer complex and in the $n=3,4$ bilayer complex.
Correspondingly, the many-body wave function can be presented as a
product of the functions (\ref{vf1}):
\begin{equation} \label{9}
\fl | \psi \rangle = \prod_X \left(\cos \frac {\theta _
{\mathrm{T} X}} {2} \, c ^ + _ {1 X} + \sin \frac {\theta _
{\mathrm{T} X}} {2} \, \rme ^ {{\rmi} \varphi _ {\mathrm{T} X}}\,
c ^ + _ {2 X} \right) \left (\cos \frac {\theta _ {\mathrm{B} X}}
{2} \, c ^ + _ {3 X} + \sin \frac {\theta _ {\mathrm{B} X}} {2}\,
\rme ^ {{\rmi} \varphi _ {\mathrm{B} X}} \, c ^ + _ {4 X} \right)
|0\rangle.
\end{equation}
In the state (\ref{9}) the energy of the system consists of three
terms:
\begin{equation} \label{9-1}
    E=E_\mathrm{T}+E_\mathrm{B}+E _ \mathrm{TB}.
\end{equation}
In (\ref{9-1}) $E_\mathrm{T} $ and $E_\mathrm{B} $ are the bilayer
energies (given by Eq. (\ref{10})). The cross term has the form
\begin{equation} \label{9-2}
    E_ \mathrm{TB} = \frac {1} {2 L_\mathrm{y}} \sum _ {X, X '}
H _ \mathrm{TB} (X-X ') \cos\theta_\mathrm{T} (X)
\cos\theta_\mathrm{B }(X '),
\end{equation}
where $\displaystyle H _\mathrm{TB} (X)=l^2 \int _ {-\infty} ^
{\infty} \rmd q \, \rme ^ {{\rmi} q X} \mathcal {H} ^ \mathrm{TB}
(q)$ with
\begin{equation} \label{65}
    \mathcal {H} ^\mathrm{TB} (q) = \frac {e^2} {2 \varepsilon l^2} \,
    \rme ^ {-d |q |} \frac {(1 - \rme ^ {-d
    | \; q |}) ^ 2} {|q |} \, \rme ^ {-\frac {q^2 l^2} {2}}.
\end{equation}
Note that in the state (\ref{9}) the cross energy does not contain
the exchange part.

The stationary homogeneous state is described by four parameters:
$\nu_i=(1+\cos \theta_{0i})/2$ and $Q_i= \rmd \varphi_{iX}/ \rmd
X$ ($i=\mathrm{T,B}$). The energy in this state  reads as
\begin{equation}\label{32-0}
    E^{(0)}=E ^ {(0)} _\mathrm{T} + E ^ {(0)} _\mathrm{B} +\frac {S} {4\pi l^2}
    \mathcal {H} ^ \mathrm{TB} (0) \cos
\theta _ {0\mathrm{T}} \cos\theta _ {0\mathrm{B}},
\end{equation}
where $E ^ {(0)} _\mathrm{T} $ ($E ^ {(0)} _\mathrm{B} $) are
determined by the equation (\ref {6-1}) with $ \theta_0 =\theta _
{0\mathrm{T}} $ ($ \theta _ {0\mathrm{B}} $) and $Q=Q_\mathrm{T}
(Q_\mathrm{B}) $.

The energy of fluctuations  is found by the same procedure as for
the bilayer system. The result is
\begin{eqnarray} \label{32-1}
E_{\rm{fl}} = \frac {S} {4\pi l^2} \sum_q \Big \{\sum _
{i=\mathrm{T, B}} \Big [\tilde{m} _ {i, z} (-q) \mathcal {K} _
{zz} ^{ii} (q) \tilde{m} _ {i, z} (q)
 + 2 \tilde{m}
_ {i, z} (-q) \mathcal {K} _ {z \varphi} ^{ii} (q)
\tilde{\varphi}_i (q) \cr + \tilde{\varphi}_i (-q) \mathcal {K} _
{\varphi \varphi} ^{ii} (q) \tilde{\varphi}_i (q) \Big] +\tilde{m}
_ {\mathrm{T}, z} (-q) \mathcal {K} _ {zz} ^ \mathrm{TB} (q)
\tilde{m} _ {\mathrm{B}, z} (q) \Big \},
\end{eqnarray}
where diagonal in $i$ components of the matrix $\mathcal {K}$ are
given by the expressions (\ref{26})-(\ref{24}), and the
non-diagonal in $i$ component is  $ \mathcal {K} ^\mathrm{TB} _
{zz} (q) = \mathcal {H} ^\mathrm{TB} (q)$.

Replacing $\tilde{m} _ {i, z} (q) $ and $ \tilde{\varphi}_i (q) $
by the operators
\begin{eqnarray} \label{37-2}
\hat {m} _ {i, z} (q) = \sqrt {\frac {2\pi l^2} {S}} \left (\frac
{\mathcal {K} _ {\varphi \varphi} ^{ii} (q)} {\mathcal {K} _ {zz}
^{ii} (q)} \right) ^ {\frac {1} {4}} ( {b} _ {i, q} +  {b} ^ + _
{i,-q}),\cr  \hat {\varphi} _i (q) = {\rm i} \sqrt {\frac {2\pi
l^2} {S}} \left (\frac {\mathcal {K} _ {zz} ^{ii} (q)} {\mathcal
{K} _ {\varphi \varphi} ^{ii} (q)} \right) ^ {\frac {1} {4}} ( {b}
_ {i, q} - {b} ^ + _ {i,-q})
\end{eqnarray}
we obtain the Hamiltonian written in terms of Bose creation and
annihilation operators
\begin{equation} \label{33}
\fl H_{\rm fl} = \sum _ {q} \Big [\sum _ {i=\mathrm{T, B}} \left
(E _ {0, i} (q) + \hbar \, q v_i\right) \left (b _ {i,\, q} ^ +
(q) b _ {i,\, q} + \frac {1} {2} \right)  + g_q (b _
{\mathrm{T},\, q} ^ + b _ {\mathrm{B},\, q}
 + b _ {\mathrm{T},\, q} b _ {\mathrm{B},-q} + h.c.)\Big]
\end{equation}
with $E _ {0, i} (q) = 2 \sqrt { \mathcal {K} _ {\varphi \varphi}
^{ii} (q) \mathcal {K} _ {zz} ^{ii} (q)}$ and
\begin{equation} \label{42}
g_q = \frac{1}{2} \mathcal {H} ^\mathrm{TB}  (q) \sqrt [4] {\frac
{\mathcal {K} _ {\varphi \varphi} ^\mathrm{TT} (q) \mathcal {K} _
{\varphi \varphi} ^\mathrm{BB} (q)} {\mathcal {K} _ {zz}
^\mathrm{TT} (q) \mathcal {K} _ {zz} ^\mathrm{BB} (q)}}.
\end{equation}
In (\ref{33}) the quantities $v_i$ are given by Eq. (\ref{40-3})
(with $Q=Q_i$ and $\nu=\nu_i$).

The Hamiltonian (\ref{33}) coincides in form with one obtained in
\cite{6} for the two-component superfluid Bose gas. Its
diagonalization
 yields
\begin{equation} \label{diag}
E =  \sum _ {q} \sum _ {\alpha=1,2} E _ {\alpha} (q)  \left ( b _
{\alpha, q} ^ + b _ {\alpha, q}+ \frac {1} {2} \right),
\end{equation}
where $b ^ +_\alpha $ ($b_\alpha $) are the operators of creation
(annihilation) of collective excitations, $E_\alpha(q)$ are the
excitation spectra. The dispersion equation for the spectra is
analogous  to one obtained \cite{10}:
\begin{equation} \label{72}
\fl [E _ {0, \mathrm{T}} ^2 (q) - (E - \hbar v_\mathrm{T} q) ^2]
[E _ {0, \mathrm{B}} ^2 (q) - (E - \hbar v_\mathrm{B} q) ^2]  - 4
g_q^2 E _ {0, \mathrm{T}} (q) E _ {0, \mathrm{B}} (q) =0.
\end{equation}

One can show (see \cite{10}) that the energies  of collective
excitations are real valued and positive if the quantities $E _
{0, j} (q) $ are real valued for all $q $, and the following
inequalities are satisfied
\begin{eqnarray} \label{72-0}
[E _ {0, \mathrm{T}} ^2 (q) - (\hbar \, {v} _\mathrm{T} q) ^2] [E
_ {0, \mathrm{B}} ^2 (q) - (\hbar \, {v} _\mathrm{B} q) ^2] - 4
g_q^2 E _ {0, \mathrm{T}} (q) E _ {0, \mathrm{B}} (q)> 0,
\\\label {72-1} E _ {0, \mathrm{T}} (q) - \hbar \, {v} _\mathrm{T} |q |> 0
\end{eqnarray}
(the condition (\ref{72-1}) can be replaced by $E _ {0,
\mathrm{B}} (q)> \hbar  {v} _\mathrm{B} q $).

\begin{figure}
\begin{center}
\includegraphics[width=10cm]{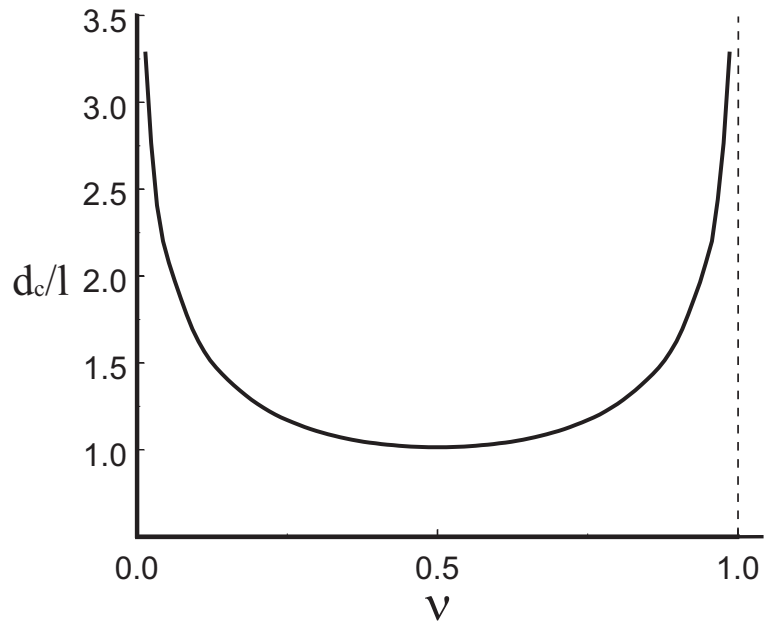}
\end{center}
\caption{Critical interlayer distance for the four-layer system vs
the filling factor ($\nu=\nu_T=\nu_B$). } \label{figure3}
\end{figure}

Putting  $Q_\mathrm{T}=Q_\mathrm{B}=0$ and solving Eq. (\ref{72})
one obtains the spectra of collective modes at zero currents. The
requirement of real valued spectra yields the critical interlayer
distance $d_\mathrm{c}$.  The dependence of $d_\mathrm{c}$ on the
filling factors is shown in Figure \ref{figure3} (for
$\nu_\mathrm{T}=\nu_\mathrm{B}$). The behavior  of the critical
interlayer distance is the same as for the bilayer system
\cite{25}, but absolute values of $d_\mathrm{c}$ are a little bit
smaller. The minimal critical distance ($d_\mathrm{c, min}\approx
1.015 l$) corresponds to the case of zero imbalance of filling
factors. At $d>d_\mathrm{c, min}$ one can say also about critical
filling factors  (that decreases under increase of $d$).

A state with nonzero supercurrents can be realized only
$d<d_\mathrm{c}(\nu_\mathrm{T},\nu_\mathrm{B})$ if for given $Q_i$
the spectra satisfy the inequalities (\ref{72-0}),(\ref{72-1}).
The currents are determined by the relation $j_i=-(e/\hbar S)\,
\rmd E^{(0)}/\rmd Q_i$. Since the cross term in Eq.(\ref{32-0})
does not depend on $Q_j$, the relation between $Q_i$ and $j_i$ is
same  (Eq. (\ref{cur})) as for the bilayer system (if one neglects
the fluctuating part of energy, see Sec. \ref{s4}). The
inequalities (\ref{72-0}),(\ref{72-1}) determine a joint condition
on $Q_i$ and the critical current of one component depends on the
current of the other component.

\begin{figure}
\begin{center}
\includegraphics[width=10cm]{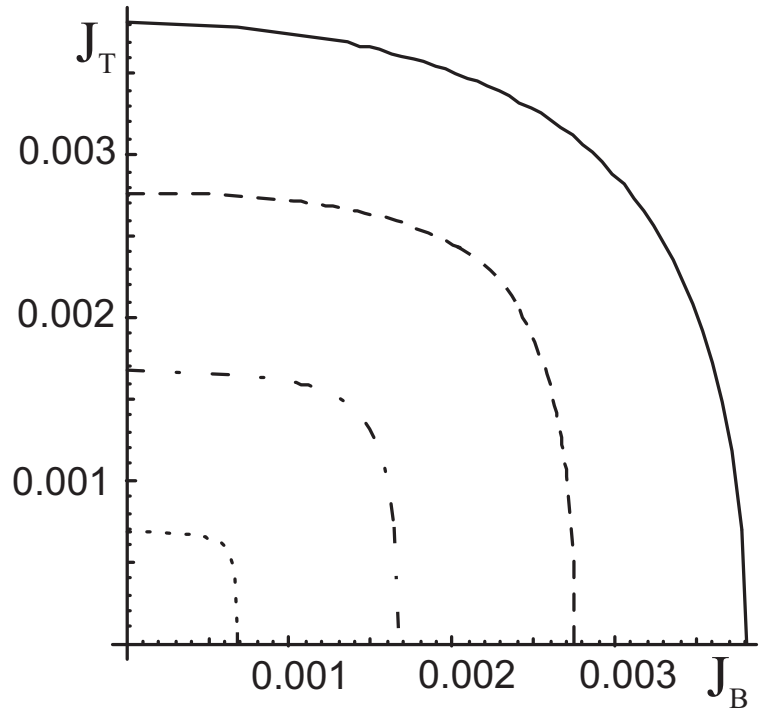}
\end{center}
\caption{Critical currents (in $e^3/\hbar \varepsilon l^2$ units)
at $d/l=0.9$ and the filling factors $\nu_T=\nu_B = 0.5, 0.25,
0.15, 0.07$ (from  the top to  the bottom curve)} \label{figure4}
\end{figure}

The relations between the critical currents for $d/l=0.9$  and
$\nu_\mathrm{T}=\nu_\mathrm{B}$ are shown in Figure \ref{figure4}.
According to Figure \ref{figure4},  typical absolute values of the
critical currents are  less or of order of 1 A/m. Comparing the
results presented in Figure \ref{figure4} with ones of Ref.
\cite{10} one can see that the critical currents demonstrate  the
behavior similar to one of critical velocities in two-component
superfluid Bose gases. Namely, the maximum supercurrent of one
component can be reached at zero supercurrent of the other
component, while at equal currents their allowed values are the
smallest one. Since measurements of electrical currents in the
layers are more simple than the measurements of superfluid
velocities in two-component mixtures,  quantum Hall four-layer
systems can be used for the observation of specific behavior of
critical velocities in two-components superfluids \cite{10}.

\section{Non-dissipative drag between the components}
\label{s4}

Eq.(\ref{cur}) used in the previous section for the calculation of
the current does not take into account the energy of fluctuations.
Therefore, the results obtained are valid at temperatures  much
smaller than the Coulomb energy. In two-component systems the
fluctuations yield an additional contribution to the current even
at zero temperatures. This contribution is caused by the energy of
zero-point oscillations. It is rather small contribution and it
can be neglected under calculations of the critical currents.
Nevertheless, this contribution determines a new effect -- a
non-dissipative drag between the components. The value of the
non-dissipative drag decreases under increase of the temperature,
but the decrease of the drag factor is essential at temperatures
larger than the interaction energy \cite{6}. Here, for simplicity,
we consider the case of zero temperatures and small phase
gradients ($Q_i l\ll 1$).

Taking into account the zero-point oscillations energy
\begin{equation}\label{1001}
    E=E^{(0)}+\frac {1} {2} \sum _ {\alpha=1,2} \sum _ {\bf q} E _
{\alpha} ({\bf q})
\end{equation}
and expanding it in series in $Q_i$ one obtains the following
expression for the energy
\begin{equation} \label{80}
 E\approx E_0 + \frac{S}{2} \sum _ {i k} \Lambda _ {i k} Q_i Q_k,
\end{equation}
where $E_0$ is independent of $Q_i$, and $\mathbf{ \Lambda}$ is
some symmetric real matrix. The mean-field energy $E^{(0)}$ in
(\ref{1001}) is diagonal in $Q_i$, but the zero-point oscillation
energy contains a non-diagonal term. Due to this a supercurrent of
a given component depends on the phase gradients of both
components
\begin{equation} \label{73}
j_i =\frac {e} {\hbar} \sum_k \Lambda_{ik} Q_k.
\end{equation}
The latter results in the non-dissipative drag effect. Indeed, let
the current  in the drive component (e.g. component T) is given
(it is fixed by an external source) while the current in the drag
component (B) is not fixed. The value of current in the drag
component  can be found from the requirement of minimum of the
energy Eq. (\ref{80}) subjected to the constrain
$j_\mathrm{T}=const$. One can see that the drag current is nonzero
and proportional to the non-diagonal component of the matrix ${\bf
\Lambda}$:
\begin{equation} \label{81}
    j_\mathrm{B} =\frac {\Lambda _\mathrm{BT}} {\Lambda _\mathrm{TT}} j _\mathrm{T}.
\end{equation}
We define  the drag factor as the ratio of the drag current to the
drive current: $f _{{\rm dr}} = {\Lambda _\mathrm{BT}}/{\Lambda
_\mathrm{TT}}$.  As was found in \cite{5,6} the drag factor for
the atomic Bose gases is rather small - the most optimistic
estimates yield  $f_{{\rm dr}}\sim 10^{-4}$. Let us compute the
drag factor for the quantum Hall multilayers.

The main contribution into the diagonal components of ${\bf
\Lambda}$ comes from the energy $E^{(0)}$, and the quantity
$\Lambda _ \mathrm{TT}$ in the leading order is evaluated as $
\Lambda _ \mathrm{TT} =\hbar^2  n_\mathrm{T}/M$
($n_i=\nu_\mathrm{i}(1-\nu_\mathrm{i})/2\pi l^2$). The
non-diagonal component of $\mathbf{\Lambda}$ is caused by the
zero-point oscillation energy. Strictly spearing, to compute this
energy one should obtain the spectrum of collective excitations
for all ${\bf q}$ (not only for ${\bf q}
\parallel \hat{x}$). But if one needs only non-diagonal in $Q_i$
term and $Q_i$ are assumed to be small an approximate dispersion
equation can be used. This equation is obtained from (\ref{72}) if
one replaces the quantities $ \hbar   {v} _i q $  with $\hbar
 {v}_i q_x $ and neglects the dependencies of $E_{0,i}$ on $Q_i$.
This approximation can be justified under accounting that in the
series for $E _ {\alpha} ({\bf q})$ the $Q_TQ_B$ terms comes only
from the product ${v}_\mathrm{B} {v}_\mathrm{T}$ (see \cite{6}).

Such  an approximation yields the dispersion equation
\begin{equation} \label{77}
[E_\mathrm{T}^2 - (E - \hbar \, {v} _\mathrm{T} q_x) ^2]
[E_\mathrm{B}^2 - (E - \hbar \, {v} _\mathrm{B} q_x) ^2] - 4
(\gamma^ {\prime})^2 \epsilon ^2 {n} _\mathrm{T} {n} _\mathrm{B} =
0,
\end{equation}
where $ E_i =\sqrt {\epsilon\big [\epsilon +2\gamma  {n} _i\big]}$
are the spectra of excitations for decoupled one-component
systems, $ \epsilon =2 \left [{\cal F} _\mathrm{D} (0) - {\cal F}
_\mathrm{D} (q) \right] $ is the kinetic energy of electron-hole
pairs,
$ \gamma=8 \pi l^2 \left [{\cal H} (q) - {\cal F} _\mathrm{S} (q)
+ {\cal F} _\mathrm{D} (0) \right] $ and $ \gamma^ {\prime} =4 \pi
l^2 \mathcal {H} ^ \mathrm{TB} (q) $ are the Fourier components of
the interaction potentials, and ${v} _i =({2}/ {\hbar \, q})
\left( {\rmd {\cal F} _\mathrm{D}(q)}/ {\rmd q}\right) (2\nu_i-1)
Q_i $ are the superfluid velocities (with the factors that account
the electron-hole symmetry).


Eq. (\ref{77}) coincides in form with one for the atomic
two-component Bose gases \cite{6,10}. For obtaining
$\Lambda_\mathrm{BT}$ we present the solutions of (\ref{77}) as
series in ${v}_i$ and substitute them into (\ref{1001}). The
details of such a procedure are described in \cite{6}. Here we
present the final expression for the drag factor
\begin{equation}\label{82}
f _\mathrm{dr}^\mathrm{BT} = \frac {2 M} {\pi \hbar^2 {n}
_\mathrm{T}} (1-2\nu_\mathrm{T}) (1-2\nu_\mathrm{B}) \int_0
^\infty \frac {(\gamma ^ {\prime}) ^2 {n} _\mathrm{T} {n}
_\mathrm{B} \epsilon^2} {E_\alpha E_\beta (E_\alpha+E_\beta)
^3}\left (\frac {\rmd \mathcal {F} _\mathrm{D}} {\rmd q} \right)^2
q \, \rmd q,
\end{equation}
where
\begin{equation} \label{18}
E _ {\alpha (\beta)} = \sqrt {\frac {E_\mathrm{T}^2 +
E_\mathrm{B}^2} {2} \pm \sqrt {\frac {(E_\mathrm{T}^2 -
E_\mathrm{B}^2) ^2} {4} + 4 (\gamma ^ {\prime}) ^2 \epsilon^2 {n}
_\mathrm{T}  {n} _\mathrm{B}}}
\end{equation}
are the energies of  collective excitations at
$Q_\mathrm{T}=Q_\mathrm{B}=0$.

\begin{figure}
\begin{center}
\includegraphics[width=10cm]{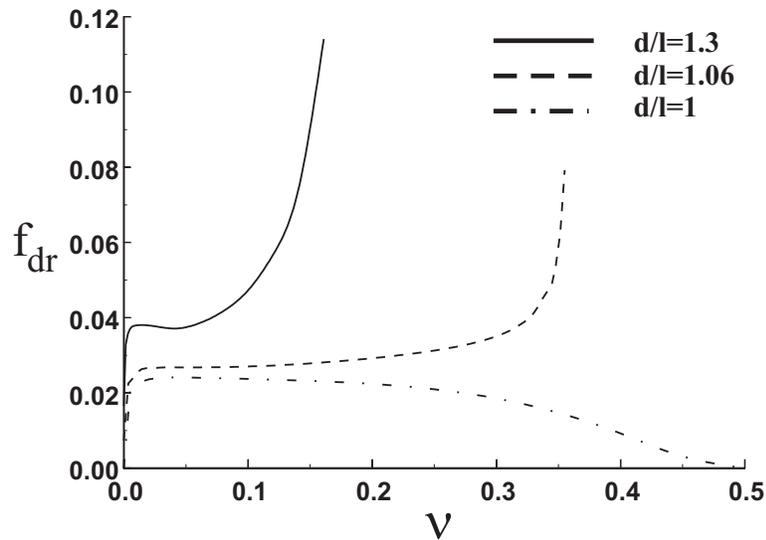}
\end{center}
\caption{Drag factor vs the filling factor ($\nu=\nu_T=\nu_B$). }
\label{figure5}
\end{figure}

Since the problem has one energy parameter (the Coulomb energy
$e^2/\varepsilon l$)  the drag factor (\ref{82}) depends only on
dimensionless quantities $d/l$, $\nu_\mathrm{T}$ and $
\nu_\mathrm{B}$. The dependence of the drag factor on the filling
factor ($\nu=\nu_\mathrm{T}=\nu_\mathrm{B}$)  at different $d/l$
is shown in Figure \ref{figure5}. One can see that this dependence
has an extremum at small $\nu$ (small density of the pairs).
Similar feature was obtained for the atomic Bose gases with
dipole-dipole interaction \cite{5}.

\begin{figure}
\begin{center}
\includegraphics[width=10cm]{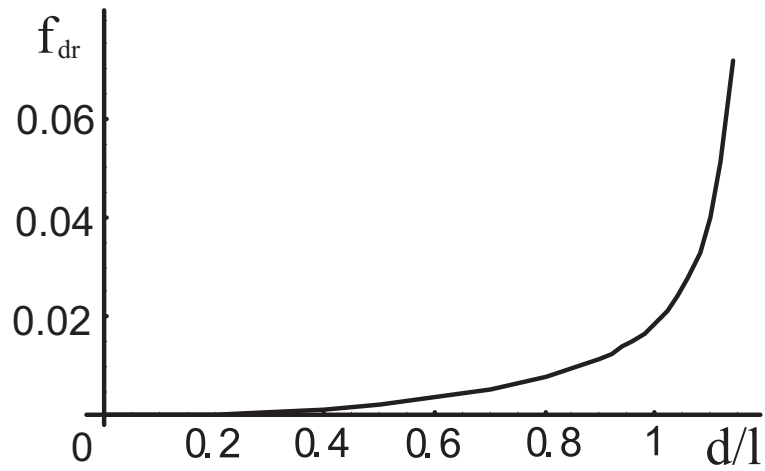}
\end{center}
\caption{Drag factor vs the interlayer distance (at
$\nu_T=\nu_B=1/4$).} \label{figure6}
\end{figure}

\begin{figure}
\begin{center}
\includegraphics[width=10cm]{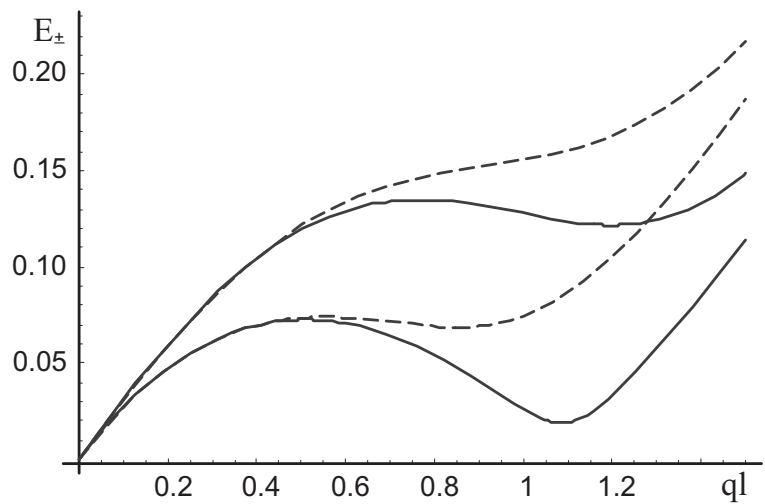}
\end{center}
\caption{The energies (in $e^2/\varepsilon l$ units) of the
collective modes  at $\nu_T=\nu_B=1/4$. Dashed curves - $d=l$,
solid curves - $d=1.16 l$} \label{figure7}
\end{figure}

The specific feature  is a sharp increase of the drag factor near
critical $\nu$ (or $d$, see Figure \ref{figure6}). The effect is
caused by a roton-like minimum in the spectrum of the lowest
collective mode (Figure \ref{figure7}).

Other specific features are the vanishing of the drag effect at
$\nu_\mathrm{B}=1/2$ ($\nu_\mathrm{T}=1/2$), and the alternation
of the sign of the drag factor under change of sign of the filling
factor imbalance ($\nu_\mathrm{T}$ to $1-\nu_\mathrm{T}$ or
$\nu_\mathrm{B}$ to $1-\nu_\mathrm{B}$). The alternation of the
direction of the drag current can be observed if one keeps one of
the filling factor constant, while the other filling factor tunes
from $\nu_i<1/2$ to $\nu_i>1/2$. This feature can be understood
from the electron-hole symmetry argumentation (see the discussion
in Sec. \ref{s2}).

But the main feature is large absolute values of the drag factor
(in $10^2\div 10^3$ times larger than the most optimistic figures
for the atomic Bose gases). Large values of the effect are caused
by a number of factors. As follows from the consideration
\cite{5,6} the drag effect in two-dimensional systems can be
larger than in three-dimensional ones, but for Bose gases in
bilayer traps large values are not reached due to  weak
interspecie interaction. Fortunately, in quantum Hall multilayers
the interaction between different superfluid components is of the
same order as the interaction inside a given component. Moreover,
the intra-component interaction is reduced due to the exchange
interaction. At last, the drag effect is enhanced considerable by
the presence of the roton-like minimum in the energy spectrum, and
this minimum becomes deeper at interlayer distances or filling
factors close to critical ones (Figure \ref{figure7}). All these
factors work in favor of the drag effect and result in a giant
drag factor in comparison with atomic two-component Bose systems.

\section{Conclusion}
In conclusion, we have studied superfluid properties of a
two-component gas of electron-hole pairs in a quantum Hall
four-layer system. We have found that the critical parameters
(critical interlayer distance and critical currents) for this
system
 are slightly less but of the same order as in bilayers. The critical
currents in the two-component gas of electron-hole pairs
demonstrate the behavior similar to one of atomic two-component
Bose gases. In particular, the largest value of the critical
current in a given component can be reached if the current in the
other component is equal to zero. In multilayer quantum Hall
systems this peculiar behavior of two-components superfluids  can
be observed by electrical measurements. The non-dissipative drag
effect between the components is predicted. The effect takes place
only at nonzero imbalance of filling factors of each (top and
bottom) bilayer, and the drag current alternates its direction
under change of sign of the imbalance in one of the bilayers. The
drag factors is quite large, and the largest values can be
achieved at the interlayer distances close to the critical ones.

\newpage

\end{document}